\documentstyle[emulateapj,onecolfloat]{article}

\slugcomment{To appear, ApJ, 1 September 2003}

\begin{document}
\twocolumn[
\title{The velocity dispersion function of early-type galaxies}
\author{Ravi K. Sheth$^1$,
Mariangela Bernardi$^2$,
Paul L. Schechter$^3$,
Scott Burles$^3$,
Daniel J. Eisenstein$^4$,
Douglas P. Finkbeiner$^{5,6,7}$,
Joshua Frieman$^8$,
Robert H. Lupton$^6$,
David J. Schlegel$^6$,  
Mark Subbarao$^8$,
K. Shimasaku$^{9}$,
Neta A. Bahcall$^6$, 
J. Brinkmann$^{10}$, and
\v{Z}eljko Ivezi\'{c}$^6$
}

\begin{abstract}
The distribution of early-type galaxy velocity dispersions, 
$\phi(\sigma)$, is measured using a sample drawn from the SDSS database.  
Its shape differs significantly from that which one obtains by simply 
using the mean correlation between luminosity $L$ and velocity 
dispersion $\sigma$ to transform the luminosity function into a 
velocity function: ignoring the scatter around the mean $\sigma-L$ 
relation is a bad approximation.  An estimate of the contribution from 
late-type galaxies is also made, which suggests that $\phi(\sigma)$ is 
dominated by early-type galaxies at velocities larger than 
$\sim 200$~km/s.  
\end{abstract}  
]

\footnotetext[1] {Department of Physics and Astronomy, 
                 University of Pittsburgh, Pittsburgh, PA 15620}
\footnotetext[2] {Department of Physics, 
                 Carnegie Mellon University, Pittsburgh, PA 15213}
\footnotetext[3] {Department of Physics, 
                  Massachusetts Institute of Technology, 
                  77 Massachusetts Avenue, Room 37-664G, 
                  Cambridge, MA 02139-4307}
\footnotetext[4] {Steward Observatory, University of Arizona, 
                  933 N. Cherry Ave., Tucson, AZ 85121}
\footnotetext[5] {Department of Astronomy, 
                  University of California at Berkeley, 
                  601 Campbell Hall, Berkeley, CA 94720}
\footnotetext[6] {Princeton University Observatory, Princeton, NJ 08544}
\footnotetext[7] {Hubble Fellow}
\footnotetext[8] {University of Chicago, Astronomy \& Astrophysics Center, 
                  5640 S. Ellis Ave., Chicago, IL 60637}
\footnotetext[9] {Department of Astronomy, University of Tokyo, 
                  Bunkyo-ku, Tokyo 113-0033, Japan}
\footnotetext[10] {Apache Point Observatory,  AZ \label{APO}}

\section{Introduction}
The distribution of internal velocities in galaxies, the 
``velocity dispersion function'', figures prominently in several 
cosmological calculations.  
Gravitational lensing cross sections (Turner, Ostriker and Gott 1984)
and the masses of central black holes (Ferrarese \& Merritt 2000; 
Gebhart et al. 2000) depend more closely on the internal velocities 
than on total mass or total light.  
The lensing efficiency of a galaxy is expected to scale as the 
fourth power of its velocity dispersion, and recent work suggests that 
the mass of the central black hole scales as the fourth or fifth power 
of the velocity dispersion of the host galaxy (Tremaine et al. 2002).  
Therefore, a reliable estimate of the velocity function is extremely 
useful.  

Perhaps more importantly, the velocity function provides a crucial 
link between models for galaxy formation and the observed universe.  
Although the shape of the mass function of dark matter halos is 
routinely predicted by theoretical models (Press \& Schechter 1974; 
Sheth \& Tormen 1999; Jenkins et al. 2001), a comparison with the mass 
function of galaxies is not straightforward.  This is because 
sufficiently massive halos host more than one galaxy, so there is 
no simple correspondence between a halo's mass and the masses of the 
galaxies it hosts.  Although galaxy formation models (White \& Rees 1978; 
White \& Frenk 1991) predict the numbers and masses of galaxies which 
form in halos as a function of parent halo mass 
(e.g., Kauffmann et al. 1999; Somerville \& Primack 1999; 
Springel et al. 2001; Benson et al. 2002), so they can be used to 
predict the mass function of galaxies, the total mass of a galaxy is 
notoriously difficult to measure (but see, e.g., Kauffmann et al. 2003 
for a technique which estimates the contribution to the mass which
comes from stars).  The same theoretical models also predict the 
distribution of luminosities, the galaxy luminosity function, which 
is much easier to measure (see, e.g., the recent determinations by 
Cross et al. 2001; Blanton et al. 2002; Madgwick et al. 2002).  However, 
this comparison between theory and observation depends, of course, on 
waveband.  The distribution of internal velocity dispersions (for 
early-type galaxies), or of circular velocities (for spiral galaxies) 
does not depend on waveband.  Since the theoretical models do predict 
$\phi(\sigma)$, the shape of the velocity function of galaxies, it is 
a more direct way of testing models of galaxy formation than is the 
luminosity function $\phi(L)$. 

Shortly after the first measurements of velocity dispersions in 
other galaxies were made (e.g., Minkowski 1954), it was recognized 
that luminosity and velocity dispersion of early-type galaxies are 
correlated (e.g., Poveda 1961; Minkowski 1962; Fish 1964; 
Faber \& Jackson 1976):  $\sigma\propto L^{1/\psi}$, 
with the exact value of $\psi$ depending on wavelength.  
And, following Tully \& Fisher (1977), a similar correlation between 
the luminosity and circular velocity $v_c$  of spiral galaxies has 
also been extensively studied (e.g., Giovanelli, Haynes et al. 1997; 
Verheijen 2000).  
Some (e.g., Shimasaku 1993; Gonzalez et al. 2000) have used these 
correlations to convert a measured distribution of luminosities 
$\phi(L)$ into an estimate of $\phi(v_c)$, simply by using the $v_c-L$ 
relation to transform variables:  $\phi(v_c)=\phi(L)\,|dL/dv_c|$.  
This procedure ignores the fact that there is scatter in the $v_c-L$ 
or $\sigma-L$ relations, and so it almost certainly underestimates the 
number of objects which have large velocity dispersions.  
Since objects with large velocity dispersions figure prominently
in a number of different arguments regarding gravitational lensing
and galaxy formation, simply changing variables is an unreliable way 
to proceed (e.g., Kochanek 1993).  

The majority of lenses are known to be early-type galaxies.  
Recently, Bernardi et al. (2003a,b,c,d) compiled a sample of 
$\sim 10^4$ early-type galaxies from the SDSS database.  
The SDSS sample includes photometric measurements in the 
$u^*$, $g^*$, $r^*$, $i^*$ and $z^*$ bands, as well as 
spectroscopic information.  
(See York et al. 2000 for a technical summary of the SDSS project; 
Stoughton et al. 2002 for a description of the Early Data Release; 
Gunn et al. 1998 for details about the camera; 
Fukugita et al. 1996, Hogg et al. 2001 and Smith et al. 2002 
for details of the photometric system and calibration; 
Lupton et al. 2001 for a discussion of the photometric data reduction 
pipeline; 
Pier et al. 2002 for the astrometric calibrations; 
Blanton et al. 2002 and Strauss et al. 2002 for details of the 
tiling algorithm and target selection.)

As Bernardi et al. (2003a) discuss, the luminosities $L$, half light radii 
$R_o$ and internal velocity dispersions $\sigma$ of the galaxies can 
all be reliably estimated from the data.  Therefore, the sample is 
well-suited to estimating the distribution of early-type galaxy 
stellar velocity dispersions.  This is done in Section~2.  
Section~3 provides a simple estimate of how the shape of the velocity 
function will be modified if the distribution of circular velocities 
from later type galaxies is added.  Since we do not have measured 
circular velocities of late-type galaxies, the results of this section 
are, perhaps, less secure.  Section~4 summarizes our results.  

\section{The distribution of early-type galaxy velocity dispersions}
The SDSS main galaxy sample from which the Bernardi et al. (2003) 
early-type sample was compiled is magnitude limited at both faint and 
bright apparent magnitudes.  In the $r^*$ band which we will use in 
this paper, $14.5\le m_{r^*}\le 17.77$.  
Therefore, the observed distribution of apparent magnitudes, sizes 
and velocities is not a fair estimate of the intrinsic distributions.  
Bernardi et al. (2003b,c) present a maximum likelihood analysis which 
accounts for the selection effects and measurement errors, at the cost 
of assuming a parametric form for the intrinsic distribution of 
luminosities $L$, sizes $R_o$ and velocity dispersions $\sigma$.  
In particular, Bernardi et al. assumed that the joint distribution of 
$M=-2.5\log_{10}L$, $\log_{10}R_o$ and $\log_{10}\sigma$ is 
tri-variate Gaussian.  
With this parameterization, their analysis showed that the 
distribution of $V\equiv\log_{10}\sigma$ is Gaussian with mean~2.20 
and rms~0.11.  

\begin{figure}[t]
 \centering
 \epsfxsize=\hsize\epsffile{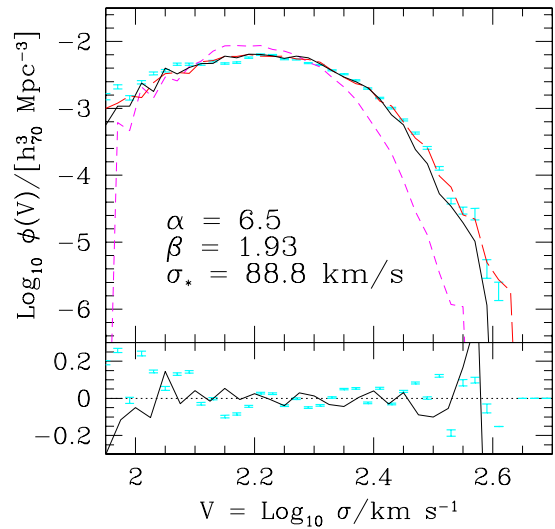}
 \caption{Velocity dispersion function in the SDSS.  
 Error bars show $\phi(\sigma)$ for the early-type galaxy sample, 
 estimated using the $1/{\cal V}_{max}$ method.  
 Short dashed line which drops most sharply shows the result of 
 transforming luminosities to velocity dispersions using the mean 
 $\sigma-L$ relation shown in the middle panel of 
 Figure~\ref{lvcontours}.  Bold solid line shows the result of 
 accounting for the scatter around the mean relation; it is well 
 described by the form given by equation~(\ref{pgamma}), with parameters 
 indicated in the panel.  Long dashed line which provides a better fit to 
 the data shows the result of convolving the intrinsic distribution 
 shown by the solid line with measurement errors.  Solid curve in 
 bottom panel shows the ratio of the solid line in the top panel 
 to the fitting formula, and error bars in bottom panel show 
 $\log_{10}$ of the ratio of the curve traced out by the error bars 
 to the long dashed line in the top panel. }
 \label{vf}
\end{figure}

Here we have chosen instead to show a nonparametric $1/{\cal V}_{max}$ 
estimate (Schmidt 1968) of the velocity dispersion function in 
Bernardi et al.'s sample.  
To calculate ${\cal V}_{max}$ we assumed a flat cosmological model 
with $\Omega=0.3$ and Hubble constant of 70 km~s$^{-1}$Mpc$^{-1}$, and 
used the SDSS apparent magnitude limits ($14.5\le m_{r^*}\le 17.77$).  
The lines with error bars in Figure~\ref{vf} show this estimate.  
The Gaussian found by Bernardi et al. provides a reasonable but not 
perfect fit, so we have not shown it.  As we discuss below, this is 
because they assumed that that the scatter around the mean $\sigma-L$ 
relation was the same for all $L$, whereas, in fact, the scatter 
depends weakly on $L$.  
The three curves in the top panel show different estimates of 
$\phi(V) = \ln(10)\,\sigma\phi(\sigma)$.  They were obtained 
as follows.  

\begin{figure}[t]
 \centering
 \epsfxsize=\hsize\epsffile{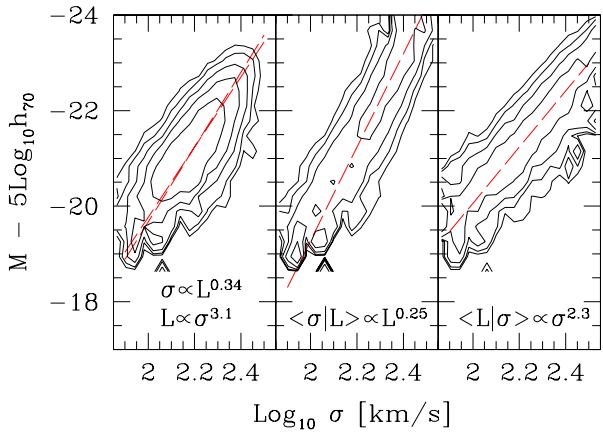}
 \vspace{-2cm}
\caption[]{Joint distribution of luminosity and velocity dispersion.  
 In each panel, contours show lines of equal probability chosen to be 
 1/2, 1/4, 1/8 etc. the value at the maximum.  
 In the panel on the left, each early-type galaxy was weighted by 
 $1/{\cal V}_{max}$, whereas the weighting was 
 $1/[\phi_e(L)\,{\cal V}_{max}]$ and $1/[\phi_e(\sigma)\,{\cal V}_{max}]$ 
 in the middle and rightmost panels.  
 The dashed line in each panel, and the text in the bottom, show 
 Bernardi et al.'s (2003b) maximum likelihood estimates of these 
 relations (sometimes called the bisector, inverse and direct fits).}
\label{lvcontours}
\end{figure}

The velocity dispersion function is 
\begin{equation}
 \phi(\sigma) = \int dL\,\phi(L)\,p(\sigma|L), 
 \label{bayes}
\end{equation}
where $p(\sigma|L)$ is the distribution of $\sigma$ at fixed $L$.  
The joint distribution of $\sigma$ and $L$ in this sample is shown 
in Figure~\ref{lvcontours}.  
In each panel, the contours show lines of equal probability, 
with levels chosen to be 1/2, 1/4, 1/8 etc. the value at the maximum. 
In the panel on the left, each galaxy was weighted by $1/{\cal V}_{max}$; 
in this case, the joint distribution $p(\sigma,L)$, is well described 
by a series of concentric ellipses whose major axes are all reasonably 
well aligned with each other.  Notice that the probability distribution 
falls-off smoothly as one moves upwards and to the right along the 
major axis of the ellipse---there is no evidence of a sharp cut-off.  
The two dashed lines show Bernardi et al.'s maximum likelihood 
estimate of the two bisector fits: the average of the $\sigma-L$ and 
$L-\sigma$ regressions given in equation~(\ref{mlfits}) below.  
The figure suggests that the distribution of velocities at fixed 
luminosity is sligthly narrower for the brightest galaxies, a 
fact we return to later.  

Since we are more interested in $p(\sigma|L)$, the distribution of 
$\sigma$ at fixed $L$, than in the joint distribution $p(\sigma,L)$, 
we weighted each galaxy by $1/[\phi(L){\cal V}_{max}]$, where 
$\phi(L)$ is the value of $\phi$ when that galaxy's luminosity is 
inserted into the luminosity function, and then re-plotted the 
joint distribution of $\sigma$ and $L$.  The middle panel of 
Figure~\ref{lvcontours} shows the resulting contour plot.  
The panel on the right shows the result of weighting each galaxy 
by $1/[\phi(\sigma){\cal V}_{max}]$; i.e., this panel shows 
$p(L|\sigma)$.  The dashed lines show 
\begin{eqnarray}
 \Bigl\langle\log_{10}\sigma\Big | M_{r*}\Bigr\rangle &=& 
  2.2 - 0.102\, (M_{r*} + 21.15 + 0.85\,z)\nonumber\\
 \Bigl\langle M_{r*}\Big | \log_{10}\sigma\Bigr\rangle &=& 
  -21.15 - 0.85\,z - 5.75\,(\log_{10}\sigma - 2.2)
 \label{mlfits}
\end{eqnarray}
from Bernardi et al. (2003b), where $z$ is the redshift.  (The values 
$-21.15$ and $2.2$ are what Bernardi et al. estimate the mean values 
of $M_{r^*}$ and $\log_{10}\sigma$ are in their sample.)  
The first of these relations is the inverse relation
($\sigma$ as a function of $L$, e.g. Schechter 1980) 
and is shown as the dashed line in the middle panel, 
whereas the second is the direct relation 
($L$ as a function of $\sigma$), and is shown in the 
panel on the right.  
The lines in the panel on the left show the bisector fits, obtained 
by averaging the direct and inverse relations.  
(If the first of the relations above is $V = (M-a_{inv})/b_{inv}$ 
and the second is $M=a_{dir} + b_{dir}V$, then the two average values 
of the slope are $M \propto V\,(b_{dir} + b_{inv})/2$ and 
$V\propto M\,(1/b_{dir}+1/b_{inv})/2$.)  
The text in the bottom of each panel gives the slopes of these 
relations.  The fact that these slopes are rather different from 
one-another is a consequence of the fact that the $\sigma-L$ relation 
is not very tight.  

When luminosities are available but velocity dispersions are not, 
equation~(\ref{bayes}) shows that $\phi(\sigma)$ can be approximated 
if one has a good model of $p(\sigma|L)$.  A first approximation, 
then is to assume that $p(\sigma|L)$ is sharply peaked about a 
characteristic value.  Because we are interested in $\sigma$ at fixed 
$L$, this characteristic value is given by the first, rather than the 
second, of equations~(\ref{mlfits}).  Comparison of the middle panel 
of Figure~\ref{lvcontours} with those on either side of it, shows that 
this choice is important:  the relations shown in the different panels 
are quite different from each other, so changing the relation used to 
transform from $L$ to $\sigma$ will change the predicted $p(\sigma)$.  
Equation~(\ref{bayes}) shows that it is the relation in the middle panel 
which should be used.  

The short dashed curve in Figure~\ref{vf} shows the result of 
transforming all observed luminosities into velocity dispersions 
$\log_{10}\sigma_{\rm FJ}$ using the first of equations~(\ref{mlfits}) 
and then making a $1/{\cal V}_{max}$ estimate of the resulting velocity 
dispersion function.  Clearly, this procedure underestimates the number 
of objects with $\log_{10}\sigma>2.3$ by a large factor.  
This shows explicitly that simply changing variables is an unreliable way 
to proceed.  We show below that this happens because the scatter around 
the mean $\langle\sigma|L\rangle$ is substantial 
(see Figure~\ref{lvcontours}).  

\begin{figure}[b]
 \centering
 \epsfxsize=\hsize\epsffile{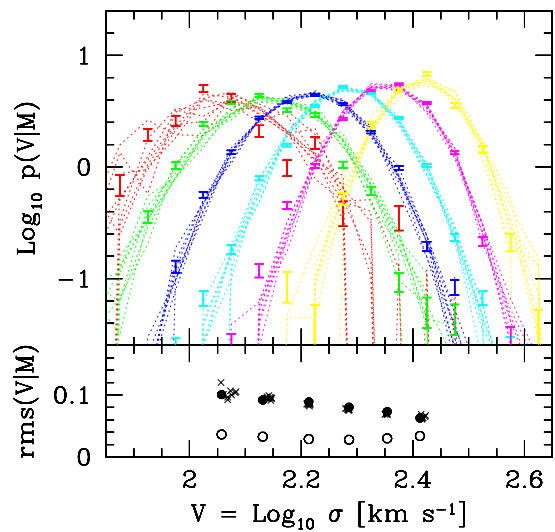}
 \caption[fz]{ Distribution of $p(V|M)$ for a number of small bins 
 in $M$.  Error bars show this distribution when the actual 
 velocity dispersions are used, and lines show the result of 
 transforming the luminosities using the mean $\sigma-L$ relation 
 and then accounting for scatter around the mean relation and 
 measurement errors.  A number of realizations of the scatter and 
 errors are shown.  Bottom panel shows the rms values; filled 
 circles show the total observed scatter, open circles show the 
 contribution from measurement errors, and crosses show the  
 total scatter in each of the realizations.}
 \label{pVMsim}
\end{figure}

To incorporate the effects of this scatter, we added a Gaussian 
random variate with 
\begin{equation}
 {\rm rms}(V|M) = 0.079\Bigl[1 + 0.17\Bigl(M_{r*}+21.15+0.85\,z\Bigr)\Bigr] 
 \label{rmsVM}
\end{equation}
to $\log_{10}\sigma_{\rm FJ}$ for each galaxy.  The solid line in 
Figure~\ref{vf} shows the resulting $1/{\cal V}_{max}$ estimate of 
$\phi(\sigma)$.  It is in much better agreement with the actual 
distribution.  The small discrepancy which remains is primarily due to 
observational errors.  To check this, we added a second Gaussian random 
variate to each $\log_{10}\sigma_{\rm FJ}$ estimate, with rms given by 
the observational error quoted in Table~2 of Bernardi et al. (2003a).  
The associated distribution is shown by the long dashed line in the top 
panel of Figure~\ref{vf}.  The error bars in the bottom panel show the 
difference between the actual measured values (indicated by the error 
bars in the top panel) and the long dashed line in the top panel.  
This indicates that our method accounts quite accurately for the effects 
of intrinsic scatter and measurement errors.  
The solid line in the bottom panel shows $\log_{10}$ of the ratio 
of the solid in the top panel to the smooth fitting function 
described in equation~(\ref{pgamma}) below.  This indicates that 
our fit is an accurate description of the intrinsic $\phi(\sigma)$ 
distribution.  

The weak link in the procedure above is the assumption that the 
distribution around the mean $\sigma-L$ relation is Gaussian 
(although the analysis in Bernardi et al. (2003b) suggests this 
should be reasonably accurate) with rms given by equation~\ref{rmsVM}.  
To check this, the error bars in Figure~\ref{pVMsim} show 
$\ln(10)\,\sigma p(\sigma|L)$ for a few narrow bins in luminosity.  
The lines show the same quantity when $\log_{10}\sigma_{\rm FJ}$ plus 
intrinsic scatter plus measurement error are used instead of the 
actual $\log_{10}\sigma$.  We have actually shown several realizations 
of this procedure, so as to give some indication of how well 
$\phi(\sigma)$ can be determined from $\sim 9000$ galaxies.  
The filled circles in the bottom panel show the measured value of the 
rms scatter for each bin in $L$, and the open circles show the typical 
contribution from measurement error (the location along the $x$ axis is 
given by the mean value of $\log_{10}\sigma$ in each bin).  The crosses 
show the estimated rms from each of the mock realizations.  Except for 
the least luminous galaxies, our procedure recovers the observed 
$p(\sigma|L)$ distribution quite well.  

\subsection{A fitting formula for $\phi(\sigma)$}
Although the velocity function can be described by transforming the 
luminosity function using the $\sigma - L$ relation and its scatter, 
it is useful to describe $\phi(\sigma)$ in a way that is independent 
of waveband.  The previous section showed that we had an accurate 
method for generating several realizations of the distribution of 
velocities.  Therefore, we made several such realizations, measured 
$\phi(\sigma)$ in each, and then found the parameters 
$(\alpha,\beta,\sigma_*)$ which provided the best fit to 
\begin{equation}
 \phi(\sigma)\,d\sigma = \phi_*\,\left({\sigma\over \sigma_*}\right)^{\alpha}\,
               {\exp[-(\sigma/\sigma_*)^{\beta}]\over\Gamma(\alpha/\beta)}\,
               \beta\,{d\sigma\over \sigma} ,
 \label{pgamma}
\end{equation}
where $\phi_*$ is the number density of galaxies.  For this sample, 
Bernardi et al. (2003b) estimate $\phi_*=0.002\,(h_{70}^{-1}$Mpc$)^{-3}$, 
whereas the $1/V_{max}$ estimate, $0.0022\,(h_{70}^{-1}$Mpc$)^{-3}$,
is slightly higher.  
We chose this functional form because, following Schechter (1976), 
the luminosity function is usually fit to such a function, but with 
$\beta$ set equal to unity.  The result of changing variables using 
$(L/L_*) = (\sigma/\sigma_*)^\psi$ would require 
$\phi(\sigma)=
 [\phi_*/\Gamma(\alpha'/\beta')]\,(\sigma/\sigma_*)^{\alpha'}\,
  \exp[-(\sigma/\sigma_*)^{\beta'}]\,\beta'/\sigma$ 
which is of the form given above.  Notice that, in this case, 
$\alpha' = \alpha\psi$ and $\beta'=\beta\psi$:  the parameters which 
specify $\phi(\sigma)$ are determined by fits to the luminosity 
function and to the $\sigma-L$ relation (e.g., Figure~\ref{lvcontours} 
shows that $\psi = 3.9$).  
Since we have already shown that simply changing variables is 
inaccurate, we chose to keep this functional form, but allowed 
$\alpha$ and $\beta$ to be determined by the fit to $\phi(\sigma)$, 
rather than by the fits to $\phi(L)$ and the $\sigma-L$ relation.  

The best fit values, 
 $(\phi_*,\sigma_*,\alpha,\beta) = 
  (0.0020\pm 0.0001, 88.8\pm 17.7, 6.5\pm 1.0, 1.93\pm 0.22)$, 
with $\phi_*$ in ($h_{70}^{-1}$Mpc)$^{-3}$ and $\sigma_*$ in km~s$^{-1}$, 
are indicated in the top panel of Figure~\ref{vf}.  
(Note that $\Gamma(6.5/1.93)\approx 2.88$.)  
The value of $\log_{10}\sigma_* = 1.95$ is, apparently, substantially 
smaller than the value 2.2 estimated by Bernardi et al. (2003b).  
This apparent discrepancy is resolved by noting that the mean value 
of $\sigma$ computed from equation~(\ref{pgamma}) is 
 $\sigma_*\,\Gamma[(\alpha+1)/\beta]/\Gamma(\alpha/\beta) \approx$ 
160~km~s$^{-1}$.  This is in excellent agreement with the mean value 
$10^{2.2} = 160$~km~s$^{-1}$ derived by Bernardi et al.  
The solid line in the bottom panel indicates that this functional form 
describes the intrinsic $\phi(\sigma)$ distribution rather well.  

\begin{figure}
 \centering
 \epsfxsize=\hsize\epsffile{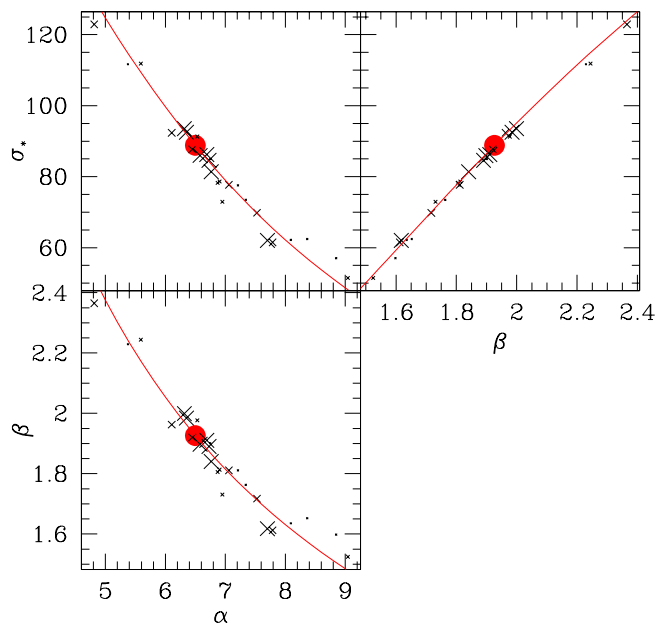}
 \caption[fz]{Covariances between the best-fit parameters obtained 
in different Monte-Carlo realizations of the velocity distribution; 
symbol size indicates goodness-of-fit.  The locii traced out by the 
symbols in each panel are well described by requiring that all fits 
yield the same values for the mean $\bar\sigma$, and the most probable 
velocity dispersion $\tilde\sigma$. Large filled circles show the 
best-fitting values we report in the previous figures.  }
\label{covar}
\end{figure}

Although we have quoted rms values around the best-fits given above, 
the best-fit values are, in fact, strongly correlated with one 
another, suggesting that the data strongly constrain some combination 
of these parameters.  Natural choices of such combinations are the 
mean, $\bar\sigma$, and most probable, $\tilde\sigma$, values of 
$\sigma$.  Equation~\ref{pgamma} shows that 
$\sigma_* = \bar\sigma\,\Gamma(\alpha/\beta)/\Gamma[(\alpha+1)/\beta]$ 
and 
$(\alpha-1)/\beta = (\tilde\sigma/\sigma_*)^\beta$, 
which allow us to provide a good approximation to the covariances 
between $\sigma_*$, $\alpha$ and $\beta$.  
This is illustrated in Figure~\ref{covar}:  symbols show best-fit 
parameter values; symbol size indicates the goodness-of-fit.  The 
smooth curves in each panel show $\beta\approx (14.75/\alpha)^{0.8}$, 
and $\sigma_* = 161\,\Gamma(\alpha/\beta)/\Gamma[(\alpha+1)/\beta]$.  
The fits return $\bar\sigma = 160\pm 1.6$~km~s$^{-1}$ values which 
only differ by about one percent, and the ratio 
$\tilde\sigma/\bar\sigma = 0.95\pm 0.007$ is also very well 
determined.  

Although we do not show it, the scatter in $\phi_*$ is slightly 
correlated with $\alpha$.  This is because if one changes the shape 
of the probability distribution by allowing more galaxies in the faint 
end (i.e., by changing $\alpha$), one must decrease the area under the 
curve at the bright end (because the area must integrate to unity).  
Since it is the bright end which is better measured, it cannot change 
too much, which means $\phi_*$ must increase to compensate.  

We have also performed the integral in equation~(\ref{bayes}) 
numerically, using the $z=0$ luminosity function from 
Bernardi et al. 2003b, and using equations~\ref{mlfits} 
and~\ref{rmsVM} for $p(\sigma|L)$.  
Fitting equation~(\ref{pgamma}) to the result gives $\alpha=8$ 
with $\sigma_*$, and $\beta$ as shown by the lines in Figure~\ref{covar}.  

The result of convolving the intrinsic distribution with measurement 
errors, modelled as a Gaussian distribution with rms 0.035 in 
$\log_{10}\sigma$, is well described by equation~(\ref{pgamma}) with 
$(\sigma_*,\alpha,\beta) = (88.8, 6.5, 1.8)$.  
The error bars in the bottom panel of Figure~\ref{vf} show $\log_{10}$ 
of the ratio of the observed $\phi(\sigma)$ to this form;    
clearly, this convolved distribution provides a good description of 
the data.  We conclude that our fitting formula, equation~(\ref{pgamma}), 
with $(\sigma_*,\alpha,\beta) = (88.8, 6.5, 1.93)$, provides a good 
description of the intrinsic $\phi(\sigma)$ distribution.

\section{The contribution from late-type galaxies}\label{lates}
So far we have studied the distribution of early-type galaxy velocity 
dispersions.  The circular velocity $v_c$ is the analogous measure of 
the potential well of a spiral galaxy.  Although we do not have 
measured values of $v_c$ for any of the SDSS galaxies which are not 
early-types, we can build a model of the contribution to the velocity 
function following the method used in the previous section:  we first 
estimate the luminosity function of galaxies which are not early 
types, $\phi_{ne}(L)$, and we then use the $v_c-L$ relation to change 
variables from $\phi_{ne}(L)$, being careful to account for inclination 
effects, which are expected to partially obscure the observed luminosities 
of late-type galaxies, and the intrinsic scatter around 
$\langle v_c|L\rangle$.  Finally, to compare with the velocity 
dispersion function of early-type galaxies, we convert from $v_c$ to 
velocity dispersion by assuming that $\sigma = v_c/\sqrt{2}$.  

We estimated $\phi_{ne}(L)$ as follows.  
The luminosity function of the entire SDSS sample, $\phi_{tot}(L)$, 
has been estimated by Blanton et al. (2003).  The SDSS photometric 
pipeline outputs a number of different estimates of the magnitude of 
a galaxy.  Although Blanton et al. use Petrosian magnitudes, 
Bernardi et al.'s (2003b) estimate of the early type galaxy 
luminosity function did not.  Therefore, we estimated the luminosity 
function of the Bernardi et al. early type sample using Petrosian 
magnitudes, $\phi_e(L)$, and then set 
$\phi_{ne}(L) = \phi_{tot}(L) - \phi_e(L)$.  
(The Petrosian luminosity accounts for about 85\% of the total 
luminosity in a deVaucouleurs profile.  We found that $\phi_e(L)$ 
was indeed well approximated by simply rescaling all luminosities 
in Bernardi et al.'s $\phi(L)$ by this factor.)

Our next problem is to estimate the $v_c-L$ relation for later-type 
galaxies.  Giovanelli et al. (1997) report that an inverse fit to the 
$v_c-L$ relation yields 
 $\log_{10}2v_c = 2.5 - 21.1/7.95 - (M_I-5\log_{10}h_{100})/7.94$.  
Applying their fitting procedure to the early type galaxy sample 
we used in the previous section yields coefficients which are not 
formally the same as those of the $\langle\sigma|L\rangle$ 
relation of equation~(\ref{mlfits}).  However, the numerical value 
of the slope only changes from $-0.102$ to $-0.104$.  Because these 
two are very close, we will assume that Giovanelli et al.'s fit 
approximates $\langle v_c|L\rangle$ closely.  

Tully et al. (1998) also report a fit to the $v_c-L$ relation, and we 
have checked that it is very close to the one from Giovanelli et al. 
which we have chosen to use.  We made this choice because  
Giovanelli et al.'s fit comes with a model for the scatter around 
the mean relation: at fixed velocity dispersion, the intrinsic scatter 
around their bivariate fit is 
 $\epsilon_{int} = 0.26 - 0.28(\log_{10}2v_c - 2.5)$~mags.  
We converted this into a scatter in $\log_{10}\sigma$ by dividing 
$\epsilon_{int}$ by $7.94$.  Notice that this makes the scatter around 
the mean $v_c-L$ relation substantially smaller than it is around 
$\langle\sigma|L\rangle$.  

\begin{figure}[b]
 \centering
 \vspace{-1cm}
 \epsfxsize=\hsize\epsffile{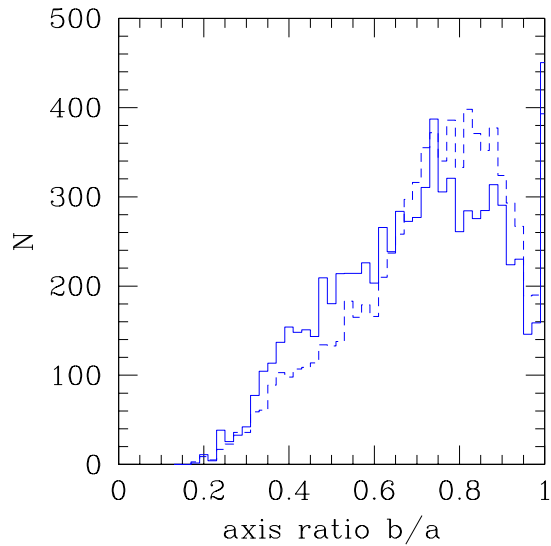}
 \caption[]{Observed distribution of early-type galaxy axis ratios 
$b/a$ (dashed line), and after weighting each galaxy by 
$1/{\cal V}_{max}$:  accounting for selection effects is 
important.  }
\label{ba}
\end{figure}

To use these results, we first converted our simulated distribution of 
$r^*$ magnitudes into $M_I^{obs}$ by setting $M_I^{obs} = M_{r^*}-0.9$ 
(this is motivated by Fukugita et al. 1995 who suggest that the 
conversion factor is 0.95 for ellipticals, 0.86 for S0s and 0.89 
for Sabs).  We then corrected luminosities to face-on values following 
the discussion in Tully et al. (1998) (also see Giovanelli et al. 1995).  
This correction makes use of the observed axis ratio $b/a$:  namely, 
\begin{eqnarray}
 M_I &=& [M_I^{obs} + g(16.9 + 5\log_{10}h_{80})]/(1-g),{\rm where}\nonumber\\ 
   g &=& -0.20\log_{10}(b/a).  
\end{eqnarray}
In practice, galaxies are observed to have a range of axis ratios.  
Khairul-Alam \& Ryden (2002) provide estimates of this distribution 
for SDSS galaxies, but we chose not to use their results because they 
do not account for selection effects.  Our concern is prompted by the 
fact that estimates of $p(b/a)$ in the Bernardi et al. sample, with and 
without $1/{\cal V}_{max}$ weighting, do differ from each other 
(see Figure~\ref{ba}).  If the intrinsic axis ratio is $r_0$, then 
\begin{equation}
 p(b/a) = (b/a)\, \sqrt{1-r_0^2\over (b/a)^2-r_0^2},
\end{equation}  
if the distribution of inclination angles is random.  
This is the distribution of $b/a$ values we chose to use.  

\begin{figure}[t]
 \centering
 \epsfxsize=\hsize\epsffile{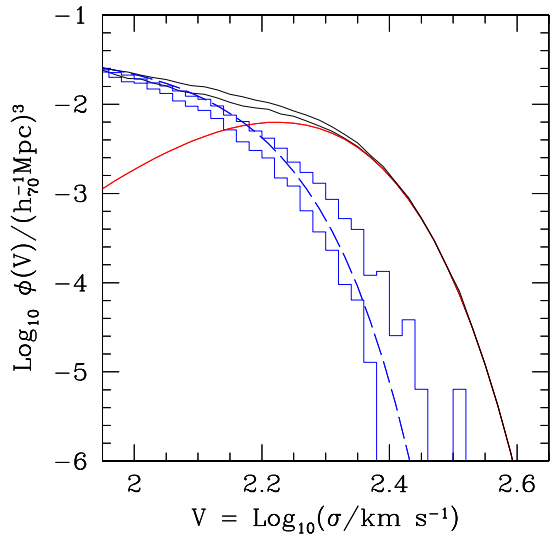}
 \caption[]{Estimate of the contribution to the distribution 
of velocities from late-type galaxies.  The two histograms show 
the effect of transforming $\phi_{ne}(L)$ and $\phi_{tot}(L)$ 
using the Tully-Fisher relation and accounting for scatter around 
the mean relation as well as correcting for inclination and 
intrinsic absorption effects.  The dashed line shows the result of 
transforming $\phi_{tot}(L)$ by ignoring the scatter around the 
Tully-Fisher relation and around the typical inclination and 
absorption correction.  The solid line shows the contribution 
to the statistic from early-type galaxies; they dominate at 
$\sigma>200$km~s$^{-1}$.  }
\label{vftot}
\end{figure}

We made mock realizations of the contribution to the velocity function 
from objects which are not early-types by assuming that all galaxies 
which are not early-types have $r_0=0.2$ and follow the $v_c-L$ 
relation above.  (In practice, there will be a range of $r_0$ values 
which our procedure ignores, but the results to follow do not depend 
strongly on the precise value of $r_0$.)   
The histogram which has the fewest galaxies with large values of 
$\sigma$ in Figure~\ref{vftot} shows this estimate.  Since the 
scatter around the mean $v_c-L$ is relatively small, accounting for 
it is not as important as it was for the early-type galaxies.  
Similarly, correcting to face-on values by assuming that all galaxies 
have the same $b/a = \langle b/a\rangle$ (i.e., setting $g=0.056$) 
and ignoring the scatter only results in a small underestimate of 
the distribution of large $v_c$ systems.  
This suggests that, for later-type galaxies, simply changing variables 
from $L$ to $v_c$ should be reasonably accurate, provided one first 
corrects all luminosities to face-on values (but see discussion below).  

The solid curve which extends furthest to the right in 
Figure~\ref{vftot} shows the contribution from early type galaxies we 
discussed in the previous section; clearly, they dominate the 
statistic at $\sigma>200$~km~s$^{-1}$.  To make this point even more 
clearly, the dashed line shows the result of assuming that all 
galaxies are spirals, and so changing variables in $\phi_{tot}(L)$ 
rather than $\phi_{ne}(L)$, and the histogram shows the effect of 
including the scatter around the mean Tully-Fisher and 
inclination/absoprtion corrections.  Our conclusion that early-types 
dominate at large velocity dispersions is still valid.  

The correction to face-on values is large, on the order of 0.5~mags, 
and, while well-defined, rather uncertain.  This is why we have not 
performed fits to the contribution from later-types, nor have we tried 
to fit equation~(\ref{pgamma}) to the sum of the two contributions 
(shown as the solid lines in Figure~\ref{vftot}).  

\section{Discussion}
We have presented estimates of the distribution of velocity dispersions 
$\phi(\sigma)$ of early-type galaxies, and have shown that estimates 
which use the mean $\sigma$ at fixed $L$, $\langle\sigma|L\rangle$, to 
change variables from $L$ to $\sigma$, and ignore the scatter around 
the mean $\sigma-L$ relation underestimate the true number density of 
large velocity dispersion systems by large factors (Figure~\ref{vf}).  
We have shown that the dependence of virial velocity dispersion on 
galaxy luminosity is a power law (Figure~\ref{lvcontours}), and we 
have derived an accurate model for the scatter in $\sigma$ at fixed 
luminosity (equation~\ref{rmsVM} and Figure~\ref{pVMsim}).  
Finally, we have provided a simple fitting formula for $\phi(\sigma)$ 
(equation~\ref{pgamma}).  

We have also built a simple model of the contribution to the velocity 
function from galaxies which are not early-types.  Our results suggest 
that, at velocity dispersions above about 200~km~s$^{-1}$, early-type 
galaxies dominate the statistics (Figure~\ref{vftot}).  Thus, we 
have demonstrated that, at large $\sigma$ the velocity dispersion 
function falls as $\exp[-(\sigma/88.8~{\rm km~s}^{-1})^{1.93}]$.  

The method we used for using observables other than $\sigma$ to 
estimate $\phi(\sigma)$ is general.  For instance, if one wishes to 
use the Fundamental Plane relation to derive $\phi(\sigma)$ from 
photometric data only, then one requires knowledge of the mean 
velocity dispersion at fixed size and surface brightness, 
$\langle\sigma|R_o,\mu_o\rangle$, as well as the scatter around this 
mean relation.  
Note that it would be incorrect to use the coefficients of the usual 
direct fit to the Fundamental Plane relation, 
$\langle R_o|\sigma,\mu_o\rangle$, (reported, e.g., in Table~2 of 
Bernardi et al. 2003c) to make the change of variables, 
for the same reason that it would have been incorrect to use the 
coefficients of the $\langle L|\sigma\rangle$ relation, rather 
than those of the $\langle\sigma|L\rangle$ relation 
(although, because the Fundamental Plane is tighter, the difference 
between the slopes will be smaller, and the effect of the scatter 
less pronounced.)

Recent work (e.g. Trujillo et al. 2001) has revived interest in 
Sersic's (1968) generalization of the deVaucouleur $(1/4)$-profile to 
$(1/n)$-profiles.  In particular, $n$ appears to be rather tightly 
correlated with $\sigma$.  Use of this correlation may be a more 
promising way (than transforming the luminosity function) to estimate 
$\phi(\sigma)$ from photometric information, and is the subject of 
work in progress.  

The mass function for clusters of galaxies cuts off sharply at large 
masses, as does the galaxy luminosity function.  Our results indicate 
that we can now add the velocity dispersion function to this list---a 
simple power law cannot describe the shape of $\phi(\sigma)$.  As 
Schechter (2002) discusses, the dearth of galaxies with large velocity 
dispsersions contains important information about the gastrophysics of 
how the most massive galaxies must have formed.  
Indeed, Loeb \& Peebles (2002) have used our measurement of $\phi(\sigma)$, 
in particular, our finding that values of $\sigma>350$~km~s$^{-1}$ are 
extremely uncommon, to argue that the stars in the galaxies with largest 
velocity dispersions must have formed at sufficiently high redshift 
that gas dissipation effects are small.  
Kochanek (2001) has pointed out that combining a lensing based 
estimate of $\phi(\sigma)$ with the one based on the motions of the 
stars, such as that presented here, provides powerful constraints on 
models of galaxy formation.  
By the time the SDSS survey is complete, a lensing based estimate 
of the velocity dispersion function should be possible.  This will 
almost certainly measure velocity dispersions on larger scales than 
the few kpc scale probed by our measurement.  Therefore, a comparison 
of the two will provide information about the effects of dissipation 
and baryonic contraction.  

\acknowledgments

PLS gratefully acknowledges the award of a John Simon Guggenheim
Fellowship and the hospitality of the Institute for Advanced Study.  

Funding for the creation and distribution of the SDSS Archive has been 
provided by the Alfred P. Sloan Foundation, the Participating
Institutions, the National Aeronautics and Space Administration, 
the National Science Foundation, the U.S. Department of Energy, 
the Japanese Monbukagakusho, and the Max Planck Society. 
The SDSS Web site is http://www.sdss.org/.

The SDSS is managed by the Astrophysical Research Consortium (ARC) for 
the Participating Institutions. The Participating Institutions are 
The University of Chicago, Fermilab, the Institute for Advanced Study, 
the Japan Participation Group, The Johns Hopkins University, 
Los Alamos National Laboratory, 
the Max-Planck-Institute for Astronomy (MPIA), 
the Max-Planck-Institute for Astrophysics (MPA), 
New Mexico State University, the University of Pittsburgh, 
Princeton University, the United States Naval Observatory, and 
the University of Washington.

{}

\end{document}